\documentclass[12pt]{article}

\usepackage[english]{babel}
\usepackage[cp1251]{inputenc}
\usepackage[T1]{fontenc}

\pdfoutput=1
\usepackage{makeidx}
\usepackage{amssymb}
\usepackage{amsfonts}
\usepackage{amsmath}
\usepackage{graphicx}
\usepackage{setspace}
\usepackage{authblk}
\usepackage{units}
\usepackage{appendix}
\usepackage{xparse}
\usepackage{cite}
\usepackage{hyperref}
\usepackage[left=2.1cm,top=2.5cm,right=2.1cm]{geometry}

\begin{document}

\title{Light beam interacting with electron medium. Exact solutions of the model and their possible applications to photon entanglement problem}
\author{A. I. Breev$^{1}$\thanks{%
		breev@mail.tsu.ru}, D. M. Gitman$^{1,2,3}$\thanks{%
		dmitrygitman@hotmail.com}, \\
	$^{1}$\small{Department of Physics, Tomsk State University, \\
		Lenin ave. 36, 634050 Tomsk, Russia;}\\
	$^{2}\ $P.N. Lebedev Physical Institute, \\
	53 Leninskiy ave., 119991 Moscow, Russia.\\
	$^{3}$ Institute of Physics, University of S\~{a}o Paulo, \\
	Rua do Mat\~{a}o, 1371, CEP 05508-090, S\~{a}o Paulo, SP, Brazil.}
\maketitle

\abstract{We consider a model for describing a QED system consisting of a photon beam
	interacting with quantized charged spinless particles. We restrict ourselves
	by a photon beam that consists of photons with two different momenta moving in
	the same direction. Photons with each moment may have two possible linear
	polarizations. The exact solutions correspond to two independent subsystems,
	one of which corresponds to the electron medium and another one is described
	by vectors in the photon Hilbert subspace and is representing a set of some
	quasi-photons that do not interact with each other. In addition, we find exact
	solution of the model that correspond to the same system placed in a constant
	magnetic field. As an example, of possible applications, we use the solutions
	of the model for calculating entanglement of the photon beam by quantized
	electron medium and by a constant magnetic field. Thus, we calculate the
	entanglement measures (the information and the Schmidt ones) of the photon
	beam as functions of the applied magnetic field and parameters of the electron medium.
	
}



\section{Introduction}\label{S1}

Exactly solvable models in quantum mechanics and QFT play important role for
understanding different physical phenomena, see e.g. Refs.
\cite{BN,BogSh80,ESM,BagGi90}. Here, we consider a model for describing a QED
system consisting of a photon beam interacting with quantized charged spinless
particles (Klein-Gordon particles that are called for simplicity electrons in
what follows). We restrict ourselves by a photon beam that consists of photons
with two different momenta $\mathbf{k}_{s}=\kappa_{s}\mathbf{n}$, $s=1,2$
(frequencies), moving in the same direction defined by a unit vector
$\mathbf{n.}$ Photons with each moment may have two possible linear
polarizations $\lambda=1,2$. In the beginning, we consider the electron
subsystem consisting of one spinless particle. Both quantized fields
(electromagnetic and the Klein-Gordon one) are placed in a box of the volume
$V=L^{3}$ and periodic conditions are supposed. For such a model we find some
exact solutions, see Sec. \ref{S2}. In these solutions we interpret
$\rho=V^{-1}$ as the electron density, believing that now the model describes
the photon beam interacting with many charged spinless particles (particles
that interact with the photon beam but do not interact with each other) the
totality of which we call below the electron medium. The exact solutions
correspond to two independent subsystems, one of which is the electron medium
and another one is described by vectors in the photon Hilbert subspace and is
representing a set of some quasi-photons that do not interact with each other.
In addition, we find exact solution of the model that correspond to the same
system placed in a constant magnetic field. As an example, we use the obtained solutions of the model for calculating entanglement of the photon beam by quantized electron medium and by a constant magnetic field, see \ref{S3}. Some final remark can be found in Sec. \ref{S4}. In the
Appendix \ref{S5} we have placed a necessary technical information, which is
used in analyzing entanglement in two-qubit systems.

\section{Photon beam interacting with quantum spinless electron
	medium}\label{S2}

\subsection{General\label{S2.1}}

We consider a model system which consists of a photon subsystem and a
subsystem of quantized charged relativistic spinless particles (the
Klein-Gordon particles). The photon subsystem under consideration (the photon
beam) consists of a finite number of photons with linear polarizations moving
in the same direction defined by a unit vector $\mathbf{n}$ (we assume that
$\mathbf{n}$ is directed along the axis $z,$ i.e., 
$\mathbf{n}=(0,0,1)$. Here, we restrict ourselves by a photon beam that consists of
photons with two different momenta $\mathbf{k}_{s}=\kappa_{s}\mathbf{n}$,
$s=1,2$ (frequencies), photons with each moment may have two possible linear
polarizations $\lambda=1,2$. We also suppose that particles from the electron
medium interact with the photon beam but do not interact with each other. Both
quantized fields (electromagnetic and the electron one) are placed in a box of
the volume $V=L^{3}$ and periodic conditions are supposed. The later
conditions imply that momenta of the photons of the beam are quantized,
\begin{equation}
	\kappa_{s}=\kappa_{0}d_{s},\quad 
	\kappa_{0}=2\pi L^{-1},\quad d_{s}\in\mathbb{N},\quad
	s=1,2\,. 
	\nonumber
\end{equation}

The operator-valued potentials $\hat{A}^{\mu}(\mathbf{r})$,
$\mathbf{r}=(x^{1},x^{2},x^{3}=z)$ of the hoton beam are
chosen in the Coulomb gauge $\partial_\mu\hat{A}^\mu=0$ as: $\hat{A}^{\mu}(\mathbf{r})
=(0,\mathbf{\hat{A}}(\mathbf{r}))$, $\operatorname{div}\mathbf{\hat{A}}(\mathbf{r})=0$. 
The nonzero vector potential depends, in fact, only on the coordinate $z$,
\begin{equation}
	\mathbf{\hat{A}}\left(  \mathbf{r}\right)  =\mathbf{\hat{A}}\left(  z\right)
	=\sum_{s=1,2}\sum_{\lambda=1,2}\sqrt{\frac{1}{2\kappa_{s}V}}\left\{  \hat
	{a}_{s\mathbf{,}\lambda}\exp\left[  i\kappa_{s}z\right]  +\hat{a}
	_{s\mathbf{,}\lambda}^{\dagger}\exp\left[  -i\kappa_{s}z\right]  \right\}
	\mathbf{e}_{\lambda}\,.
	\nonumber
\end{equation}
Here $\hat{a}_{s,\lambda}$ and $\hat{a}_{s,\lambda}^{\dagger}$ are free
photon creation and annihilation operators satisfying Bose-type commutation
relations:
\begin{equation}
	\left[  \hat{a}_{s\mathbf{,}\lambda},\hat{a}_{s^{\prime},\lambda^{\prime}
	}\right]  =0,\ \left[  \hat{a}_{s\mathbf{,}\lambda},\hat{a}_{s^{\prime
		},\lambda^{\prime}}^{\dagger}\right]  =\delta_{s,s^{\prime}}\delta
	_{\lambda,\lambda^{\prime}},\ s,s^{\prime}=1,2,\ \ \lambda,\lambda^{\prime
	}=1,2\,,d
	\nonumber
\end{equation}
where $\mathbf{e}_{\lambda}$ are real polarization vectors, 
$(\mathbf{e}_{\lambda}\mathbf{e}_{\lambda^{\prime}})=\delta_{\lambda,\lambda^{\prime}}$,
$(\mathbf{ne}_{\lambda})=0$. The photon operators are acting in a photon Fock
space $\mathfrak{H}_{\mathrm{\gamma}}$ constructing by the creation and
annihilation operators and a vacuum vector $\left\vert 0\right\rangle_{\mathrm{\gamma}}$, 
$\hat{a}_{s,\lambda}\left\vert 0\right\rangle_{\mathrm{\gamma}}=0$, 
$\forall s,\lambda$. Vectors from the photon Fock
space are denoted by $\left\vert \Psi\right\rangle _{\mathrm{\gamma}}$, $\left\vert \Psi\right\rangle _{\mathrm{\gamma}}\in\mathfrak{H}_{\mathrm{\gamma}}$.

The free photon Hamiltonian has the form:
\begin{equation}
	\hat{H}_{\mathrm{\gamma}}=\sum_{s=1,2}\sum_{\lambda=1,2}\kappa_{s}\hat{a}%
	_{s,\lambda}^{\dagger}\hat{a}_{s,\lambda}\,. 
	\nonumber
\end{equation}

On the quantum level, electrons are described by a scalar field 
$\varphi(\mathbf{r})$ (the Klein-Gordon field) interacting with a
classical external electromagnetic field of the form 
$A_{\mathrm{ext}}^{\mu}(\mathbf{r})=(A_{\mathrm{ext}}^{0}=0,\mathbf{A}_{\mathrm{ext}}(r))$, $\mathbf{A_{\mathrm{ext}}(r})=(A\mathbf{_{\mathrm{ext}}}^{k}(\mathbf{r}))$, 
$\mu=0,\dots,3$, $k=1,2,3$. We suppose that this field
does not violate the vacuum stability, see \cite{FraGiS91,GavGi80}. After
canonical quantization (see \cite{Schwe61,GavGi80}) the scalar field and its
canonical momentum $\pi(\mathbf{r})$ become operators $\hat{\varphi}(\mathbf{r})$ and $\hat{\pi}(\mathbf{r}).$ The corresponding Heisenberg operators $\hat{\varphi}(x)$ and $\hat{\pi}(x)$, $x=(x^{\mu})=(x^{0}=t,\mathbf{r})$, satisfy equal-time nonzero commutation relations:
\begin{equation}
	\lbrack\hat{\varphi}(x),\hat{\pi}(x^{\prime})]_{t=t^{\prime}}=
	i\delta(\mathbf{r}-\mathbf{r}^{\prime})\,. 
	\label{1.2c}
\end{equation}
The scalar field operators are acting in an electron Fock space $\mathfrak{H}_{\mathrm{e}}$ constructed by a set of creation and annihilation operators of scalar particles interacting 
with the external field and by a corresponding vacuum vector 
$\left\vert 0\right\rangle_{\mathrm{e}}$, see e.g. the two
latter references. Vectors from the electron Fock space are denoted as $\left\vert \Psi\right\rangle_{\mathrm{e}}\subset\mathfrak{H}_{\mathrm{e}}$. The Fock space $\mathfrak{H}$
of the complete system is a tensor product of the photon Fock space and the
electron Fock space, 
$\mathfrak{H=H}_{\mathrm{\gamma}}\otimes\mathfrak{H}_{\mathrm{e}}$. Vectors from the Fock space $\mathfrak{H}$ are denoted by $\left\vert\Psi\right\rangle$ such that, 
$\left\vert \Psi\right\rangle\in\mathfrak{H}$.

The Hamiltonian of the complete system has the following form (see, e.g. \cite{GavGi80}):
\begin{eqnarray}
	  \hat{H}&=&\int\left\{  \hat{\pi}^{+}(\mathbf{r})\hat{\pi}(\mathbf{r}
	)+\hat{\varphi}^{+}\left(  \mathbf{r}\right)  \left[  \mathbf{\hat{P}
	}_{\mathrm{\gamma}}^{2}(\mathbf{r})+m^{2}\right]  \hat{\varphi}\left(
	\mathbf{r}\right)  \right\}  d\mathbf{r}+\hat{H}_{\mathrm{\gamma}
	}\ ,\nonumber\\
	\mathbf{\hat{P}}_{\mathrm{\gamma}}(\mathbf{r})&=&\mathbf{\hat{p}}+e\left[
	\mathbf{\hat{A}(}z)+\mathbf{A_{\mathrm{ext}}(r})\right]\,.
	\label{sc1b}
\end{eqnarray}
Here $m$ is the electron mass and $e>0$ is the absolute value of the electron charge.

Consider the amplitude
\begin{equation}
	\varphi_{\mathrm{\gamma}}(x)=\ _{\mathrm{e}}\left\langle 0\right\vert
	\hat{\varphi}(\mathbf{r})\left\vert \boldsymbol{\Psi}\left(t\right)
	\right\rangle,\, 
	\label{1.12}
\end{equation}
which is the projection of the vector $\left\vert\mathbf{\Psi}(t)\right\rangle$ onto a one-electron state remaining at the same time an abstract vector in the photon Fock space\footnote{
	Of course, to be consistent, we should have used a notation like $\left\vert\Psi_{\mathrm{\gamma}}(x)\right\rangle$ for the vector. However, here and in what follows, we denote this vector as $\varphi_{\mathrm{\gamma}}(x)$, to simplify notations.
}. We interpret 
$\varphi_{\mathrm{\gamma}}(x)$ as a state vector of photons interacting with an
electron. In similar manner, one can introduce many-electron or positron
amplitudes and interpreted them as state vectors of photons interacting with
many charged particles.

To derive an equation for the state vector (\ref{1.12}) we take into account
that has the vector $\left\vert\mathbf{\Psi}(t)\right\rangle$
satisfies the Schr\"{o}dinger equation with the Hamiltonian (\ref{sc1b}),
\begin{equation}
	i\partial_{t}\left\vert\boldsymbol{\Psi}\left(t\right)\right\rangle
	=\hat{H}\left\vert\boldsymbol{\Psi}\left(t\right)\right\rangle\,.
	\label{1.13}
\end{equation}
Then with account taken of both Eqs. (\ref{1.12}) and (\ref{1.13}) and the
commutation relations (\ref{1.2c}), we obtain:
\begin{eqnarray}
	  \ i\partial_{t}\varphi_{\mathrm{\gamma}}(x)&=&\ _{\mathrm{e}}\left\langle
	0\right\vert \hat{\varphi}(\mathbf{r})\hat{H}\left\vert \boldsymbol{\Psi
	}\left(  t\right)  \right\rangle =\ _{\mathrm{e}}\left\langle 0\right\vert
	\hat{H}\hat{\varphi}(\mathbf{r})+\left[  \hat{\varphi}(\mathbf{r}),\hat
	{H}\right]  \left\vert \boldsymbol{\Psi}\left(  t\right)  \right\rangle
	\nonumber\\
	  \ &=&\chi_{\mathrm{\gamma}}(x)+\hat{H}_{\mathrm{\gamma}}\varphi
	_{\mathrm{\gamma}}(x)+\Delta\ ,\ \ \chi_{\mathrm{\gamma}}(x)=i\ _{\mathrm{e}
	}\left\langle 0\right\vert \hat{\pi}^{+}(\mathbf{r})\left\vert
	\boldsymbol{\Psi}\left(  t\right)  \right\rangle \,, 
	\label{1.13c}
\end{eqnarray}
where the vector $\chi_{\mathrm{\gamma}}(x)$ satisfies the following equation
\begin{eqnarray}
	  \ i\partial_{t}\chi_{\mathrm{\gamma}}(x)&=&i\ _{\mathrm{e}}\left\langle
	0\right\vert \hat{\pi}^{+}(\mathbf{r})\hat{H}\left\vert \boldsymbol{\Psi
	}\left(  t\right)  \right\rangle =i\ _{\mathrm{e}}\left\langle 0\right\vert
	\hat{H}\hat{\pi}^{+}(\mathbf{r})+\left[  \hat{\pi}^{+}(\mathbf{r}),\hat
	{H}\right]  \left\vert \boldsymbol{\Psi}\left(  t\right)  \right\rangle
	\nonumber\\
	  &=&\left[  \mathbf{\hat{P}}_{\mathrm{\gamma}}^{2}(\mathbf{r})+m^{2}\right]
	\varphi_{\mathrm{\gamma}}(x)+\hat{H}_{\mathrm{\gamma}}\chi_{\mathrm{\gamma}
	}(x)+\Delta\,,
	\label{1.13e}
\end{eqnarray}
and the vector $\Delta$ contains many-particle amplitudes that are related to
processes of virtual pair creation. We note that the vector $\Delta$ contains
higher-order contributions in the fine structure constant. That is why, we
neglect this term in what follows. With $\Delta=0$ equations (\ref{1.13c}) and
(\ref{1.13e}) are reduced to a set of two coupled equations for the vectors
$\chi_{\mathrm{\gamma}}(x)$ and $\varphi_{\mathrm{\gamma}}(x)$. This set
implies the following second-order equation for the vector 
$\varphi_{\mathrm{\gamma}}(x)$:
\begin{equation}
	\left[  (i\partial_{t}-\hat{H}_{\mathrm{\gamma}})^{2}-\mathbf{\hat{P}
	}_{\mathrm{\gamma}}^{2}(\mathbf{r})-m^{2}\right]  \varphi_{\mathrm{\gamma}
	}(x)=0\,. 
	\label{1.13d}
\end{equation}
It is convenient to pass from the vector $\varphi_{\mathrm{\gamma}}(x)$ to a vector $\Phi_{\mathrm{\gamma}}(x)$,
\begin{equation}
	\Phi_{\mathrm{\gamma}}(x)=\exp\left(  i\hat{H}_{\mathrm{\gamma}}t\right)
	\varphi_{\mathrm{\gamma}}(x)\,.
	\nonumber
\end{equation}
As it follows from Eq. (\ref{1.13d}) the vector $\Phi_{\mathrm{\gamma}}(x)$
satisfies a Klein-Gordon like equation:
\begin{align}
	&  \left(  \hat{P}_{\mu}\hat{P}^{\mu}-m^{2}\right)  \Phi_{\mathrm{\gamma}
	}(x)=0\ ,\label{1.14}\\
	&  \hat{P}^{\mu}=i\partial^{\mu}+e\left[  \hat{A}^{\mu}(u)+A_{\mathrm{ext}
	}^{\mu}(\mathbf{r})\right]  \ ,\ \hat{A}^{\mu}(u)=\left(  0,\mathbf{\hat{A}
	}(u)\right)  \ ,\nonumber\\
	&  \mathbf{\hat{A}}(u)=\frac{1}{e}\sum_{s=1,2}\sum_{\lambda=1,2}\sqrt
	{\frac{\varepsilon}{2\kappa_{s}}}\left[  \hat{a}_{s,\lambda}\exp\left(
	-i\kappa_{s}u\right)  +\hat{a}_{s,\lambda}^{\dagger}\exp\left(  i\kappa
	_{s}u\right)  \right]  \mathbf{e}_{\lambda}\ ,\nonumber\\
	&  \ u=t-z,\ \partial_{\mu}=\left(  \partial_{t},\mathbf{\nabla}\right)
	,\ \varepsilon=\alpha\rho,\ \alpha=e^{2}/\hbar c=1/137\,.\nonumber
\end{align}

As was already said, we interpret $\rho=V^{-1}=L^{-3}$ as the electron media
density. The quantity $\varepsilon$ characterizes the strength of the
interaction between the quantum charged particles and the photon beam.

\subsection{The system in the absence of an external field\label{S2.2}}

First we consider the case without the external electromagnetic field,
$A_{\mathrm{ext}}^{\mu}(\mathbf{r})=0$. Then equation (\ref{1.14}) admits the
following integrals of motion\footnote{Recall that an operator $\hat{G}$ is
	called an integral of motion if its mean value
	\begin{equation}
		\left(  \varphi,\hat{G}\varphi\right)  =\int\varphi^{\ast}\left(  x\right)
		\left(  i\overleftrightarrow{\partial_{0}}-2eA_{0}\right)  \hat{G}
		\varphi\left(  x\right)  d\mathbf{r\,},\ \overleftrightarrow{\partial_{0}}=
		\overrightarrow{\partial_{0}}-\overleftarrow{\partial_{0}}\,,
		\nonumber
	\end{equation}
	calculated using any wave function $\varphi$ satisfying the Klein-Gordon
	equation does not depend on time. For the operator $\hat{G}$ to be an integral
	of motion, it is sufficient that it commutes with the Klein-Gordon operator
	$\hat{P}_{\mu}\hat{P}^{\mu}-m^{2}$. If the operator $\hat{G}$ is an integral
	of motion, one can impose the condition that, apart from satisfying the
	Klein-Gordon equation, the wave function should be an eigenfunction of
	$\hat{G}$.
	\par
	{}}:
\begin{align}
	&  \left[  \hat{G}_{\nu},\left(  \hat{P}_{\mu}\hat{P}^{\mu}-m^{2}\right)
	\right]  =0,\ \ \hat{G}_{\nu}=i\partial_{\nu}+n_{\nu}\hat{H}_{\mathrm{\gamma}
	},\ \ n^{\mu}=(1,0,0,1)\ ,\nonumber\\
	&  [\hat{G}_{\mu},\hat{G}_{\nu}]=0\ ,\ \ [\hat{G}_{\mu},\hat{P}_{\nu}
	]=0,\ \mu,\nu=0,1,2,3\,. 
	\label{5.2}
\end{align}
In this case, we look for solutions of Eq. (\ref{1.14}) that are eigenvectors
for the operators $\hat{G}_{\mu}$,
\begin{equation}
	\hat{G}_{\mu}\Phi_{\mathrm{\gamma}}(x)=g_{\mu}\Phi_{\mathrm{\gamma}}(x)\,.
	\label{5.3}
\end{equation}
With account taken of Eqs. (\ref{5.3}) such solutions $\Phi_{\mathrm{\gamma}}(x)$ satisfy the following equation:
\begin{equation}
	\left[  \hat{H}_{\mathrm{K}}-\frac{g^{2}-m^{2}}{2(ng)}\right]  \Phi
	_{\mathrm{\gamma}}(x)=0,\ \ \hat{H}_{\mathrm{K}}=\hat{H}_{\mathrm{\gamma}}+
	\frac{e\left(  \mathbf{p\hat{A}}\right)  }{\left(  ng\right)  }+\frac
	{e^{2}\mathbf{\hat{A}}^{2}}{2\left(  ng\right)  }\,,
	\nonumber
\end{equation}
where $ng=g_{0}+g_{3}$.
The latter fact implies that the operator $\hat{H}_{\mathrm{K}}$ commutes with
the one $\hat{P}_{\mu}\hat{P}^{\mu}-m$ on solutions, which means:
\[
	\left[  \hat{H}_{\mathrm{K}},\left(  \hat{P}_{\mu}\hat{P}^{\mu}-m^{2}\right)
	\right]  \Phi_{\mathrm{\gamma}}(x)=0\,,
\]
and therefore the operator $\hat{H}_{\mathrm{K}}$ is also an integral of
motion. The operator $\hat{G}_{\mu}\,$can be represented in the form:
\begin{equation}
	\hat{G}_{\mu}=(\hat{P}_{\mathrm{K}})_{\mu}+n_{\mu}\hat{H}_{\mathrm{K}
	},\ \ (\hat{P}_{\mathrm{K}})_{\mu}=i\partial_{\mu}-n_{\mu}\left(  \hat
	{H}_{\mathrm{K}}-\hat{H}_{\mathrm{\gamma}}\right)\,. 
	\label{5.5}
\end{equation}
Since the operator $(\hat{P}_{\mathrm{K}})_{\mu}$ is a linear combinations of
the integrals of motion $\hat{G}_{\mu}$ and $\hat{H}_{\mathrm{K}}$, it also is
an integral of motion.

Thus, we may find solutions of Eqs. (\ref{1.14}) and (\ref{5.3}), that are at
the same time eigenvectors of the integrals of motion $(\hat{P}_{\mathrm{K}})_{\mu}$ and $\hat{H}_{\mathrm{K}}$:
\begin{eqnarray}
	  (\hat{P}_{\mathrm{K}})_{\mu}\ \Phi_{\mathrm{\gamma}}(x)&=&p_{\mu}
	\Phi_{\mathrm{\gamma}}(x)\ ,\nonumber\\ 
	  \hat{H}_{\mathrm{K}}\Phi_{\mathrm{\gamma}}(x)&=&E_{\mathrm{ph}}
	\Phi_{\mathrm{\gamma}}(x)\,. 
	\label{5.6b}
\end{eqnarray}
This consideration becomes consistent if $p^{2}=p_{0}^{2}+\mathbf{p}^{2}=m^{2}$. Without loss of generality, we can set $\mathbf{p}_{\perp}=(p^{1},p^{2})=0$ in the case under consideration. Then $(ng)=(np)=p_{0}+p_{3}$. Besides, it follows from the above equations that
\begin{equation}
	g_{\mu}=p_{\mu}+n_{\mu}E_{\mathrm{ph}}\Longrightarrow g_{0}=p_{0}
	+E_{\mathrm{ph}},\ \ g_{3}=p_{3}-E_{\mathrm{ph}}\,. 
	\label{5.7}
\end{equation}

The operator $\hat{G}^{\mu}$ can be interpreted as a four-momentum operator of
an electron and the photon beam and its eigenvalues $g_{\mu}$ as the
energy-momentum of such a system. Switching off the electron-photon
interaction, $\hat{G}^{\mu}$ is reduced to the sum of the energy momentum
operator $i\partial_{\mu}$ of the electron and of the energy-momentum operator
of the photon beam $n_{\mu}\hat{H}_{\mathrm{\gamma}}$. Thus, the system splits
into two independent subsystems - a system of quasi-photons and a system of
electrons, so that the energy-momentum vector of the system $g^{\mu}$ is the
sum (\ref{5.7}) of the energy-momentum $p^{\mu}$ of an electron and the
momentum energy of the quasi-photons $n^{\mu}E_{\mathrm{ph}}$.

Solving Eq. (\ref{5.3}), we represent the vector $\Phi_{\mathrm{\gamma}}(x)$
in the following form:
\begin{equation}
	\Phi_{\mathrm{\gamma}}(x)=\hat{U}\Phi_{\mathrm{\gamma}}^{\left(  1\right)
	}(x),\ \ \hat{U}=\exp\left(  i\hat{H}_{\mathrm{\gamma}}u\right)\,.
	\label{5.9}
\end{equation}
Since $\hat{U}^{\dagger}\hat{G}_{\mu}\hat{U}=i\partial_{\mu}$, we obtain:
\begin{equation}
	i\partial_{\mu}\Phi_{\mathrm{\gamma}}^{\left(  1\right)  }(x)=g_{\mu}
	\Phi_{\mathrm{\gamma}}^{\left(  1\right)  }(x)\Longrightarrow\Phi
	_{\mathrm{\gamma}}^{\left(  1\right)  }(x)=\exp\left(-igx\right)
	\Phi_{\mathrm{ph}}\,, 
	\nonumber
\end{equation}
where the vector $\Phi_{\mathrm{ph}}\in\mathfrak{H}_{\mathrm{\gamma}}$ does
not depend on $x$.

With account taken of Eqs. (\ref{5.9}) and (\ref{5.6b}) as well as the
operator relation $\hat{Q}^{\mu}=eU^{\dagger}\hat{A}^{\mu}\left(u\right)
U=e\hat{A}^{\mu}\left(  0\right)  =(0,\mathbf{\hat{Q}})$, one
can verify that the vector $\Phi_{\mathrm{ph}}$ satisfies the stationary
Schr\"{o}dinger equation:
\begin{equation}
	\hat{H}_{\mathrm{ph}}\Phi_{\mathrm{ph}}=E_{\mathrm{ph}}\Phi_{\mathrm{ph}
	}\ ,\ \hat{H}_{\mathrm{ph}}=\hat{U}^{\dagger}\hat{H}_{\mathrm{K}}\hat{U}
	=\hat{H}_{\mathrm{\gamma}}+\frac{\mathbf{\hat{Q}}^{2}}{2\left(np\right)
	}\,. 
	\label{5.12b}
\end{equation}

Let us perform a linear canonical transformation from the free photon creation
and annihilation operators $\hat{a}^{\dagger}$ and $\hat{a}$ to new creation
and annihilation operators $\hat{c}^{\dagger}$ and $\hat{c}$,
\begin{equation}
	\hat{a}=u\hat{c}-v\hat{c}^{\dagger},\ \ \hat{a}^{\dagger}=\hat{c}^{\dagger
	}u^{\dagger}-\hat{c}v^{\dagger},\ [\hat{c}_{s,\lambda},\hat{c}_{s^{\prime
		},\lambda^{\prime}}^{\dagger}]=\delta_{s,s^{\prime}}\delta_{\lambda
		,\lambda^{\prime}}\,, 
	\label{5.13}
\end{equation}
where $uu^{\dagger}-vv^{\dagger}=1$ and $vu^{T}-uv^{T}=0$. We call the
operators $\hat{c}^{\dagger}$ and $\hat{c}$ quasi-photon operators in what follows.

First of all, we choose the matrices $u=(u_{s,\lambda;s^{\prime},\lambda^{\prime}})$, $v=(v_{s,\lambda;s^{\prime},\lambda^{\prime}})$ to
vanish all the terms linear in the quasi-photon operators in the Hamiltonian
$\hat{H}_{\mathrm{ph}}$. Such matrices $u$ and $v$ and the column $\gamma$
have the form:
\begin{align}
	&  u_{s,\lambda;s^{\prime},\lambda^{\prime}}=u_{s,s^{\prime}}\delta
	_{\lambda,\lambda^{\prime}},\ \ v_{s,\lambda;s^{\prime},\lambda^{\prime}
	}=v_{s,s^{\prime}}\delta_{\lambda,\lambda^{\prime}}\ ,\nonumber\\
	&  u_{s,s^{\prime}}=F_{s,s^{\prime}}^{+}q_{s,s^{\prime}},\ \ v_{s,s^{\prime}
	}=F_{s,s^{\prime}}^{-}q_{s,s^{\prime}},\ \ F_{s,s^{\prime}}^{\pm}=\frac{1}
	{2}\left(  \sqrt{\frac{\tau_{s^{\prime}}}{\kappa_{s}}}\pm\sqrt{\frac
		{\kappa_{s}}{\tau_{s^{\prime}}}}\right)  \ ,\nonumber\\
	&  q_{s,s^{\prime}}=\frac{q_{s^{\prime}}}{\tau_{s^{\prime}}^{2}-\kappa_{s}
		^{2}},\ \ q_{s}\sqrt{\sum_{l=1,2}\left(  \tau_{s}^{2}-\kappa_{l}^{2}\right)
		^{-2}}=1,\quad s=1,2\,, 
	\label{1.25}
\end{align}
where the quantities $\tau_{s}$ are roots of the algebraic equation
\begin{equation}
	\epsilon\sum_{l=1,2}\left(  \tau_{s}^{2}-\kappa_{l}^{2}\right)^{-1}
	=1,\ \ \epsilon=\varepsilon/\left(  np\right)\,. 
	\label{1.27}
\end{equation}
Note that there is no summation over the repeated indices in Eqs. (\ref{1.25}).

Hamiltonian (\ref{5.12b}) is diagonal with respect to the operators $\hat{a}_{s\mathbf{,}\lambda}^{\dagger}$ and $\hat{a}_{s\mathbf{,}\lambda}$ at
$\epsilon=0$. That is why we chose roots of equation (\ref{1.27}) to satisfy
the conditions $r_{s}^{2}(\epsilon=0)=\kappa_{s}^{2}$. Thus, we obtain:
\begin{align}
	\tau_{1}^{2} &  =A+B,\ r_{2}^{2}=A-B,\ A=\epsilon+\frac{\kappa_{1}^{2}
		+\kappa_{2}^{2}}{2}\ ,\nonumber\\
	\ B &  =\mathrm{sgn}(\kappa_{1}-\kappa_{2})\sqrt{\epsilon^{2}+\left(
		\frac{\kappa_{1}^{2}-\kappa_{2}^{2}}{2}\right)  ^{2}}\,,
	\nonumber
\end{align}
where $\mathrm{sgn}(x)=1$, $x\geq0$, and $\mathrm{sgn}(x)=-1$, $x<0$.

In terms of the quasi-photon operators, Hamiltonian (\ref{5.12b}) takes the
form:
\begin{equation}
	\hat{H}_{\mathrm{ph}}=\sum_{s=1,2}\sum_{\lambda=1,2}\tau_{s}\hat{c}
	_{s,\lambda}^{\dagger}\hat{c}_{s,\lambda}+H_{0},\ H_{0}=\sum_{s=1,2}\left(
	\tau_{s}-\kappa_{s}\right)  ,\ \ \tau_{s}>0\,. 
	\nonumber
\end{equation}

The eigenvalue problem (\ref{5.12b}) has the following solution:
\begin{eqnarray}
	  E_{\mathrm{ph}}&=&\sum_{s=1,2}\sum_{\lambda=1,2}\tau_{s}N_{s,\lambda}
	+H_{0}\ ,\ N_{s,\lambda}\in\mathbb{N}\ ,\nonumber\\
	  \left\vert \Phi_{\mathrm{ph}}\right\rangle &=&\left[  \prod_{\lambda
		=1,2}\frac{\left(  \hat{c}_{1,\lambda}^{\dagger}\right)  ^{N_{1,\lambda}}
	}{\sqrt{N_{1,\lambda}!}}\left\vert 0_{1}\right\rangle _{c}\right]
	\otimes\left[  \prod_{\lambda^{\prime}=1,2}\frac{\left(  \hat{c}
		_{2,\lambda^{\prime}}^{\dagger}\right)  ^{N_{2,\lambda^{\prime}}}}
	{\sqrt{N_{2,\lambda^{\prime}}!}}\left\vert 0_{2}\right\rangle _{c}\right]\,,
	\label{1.28a}
\end{eqnarray}
where $\left\vert 0_{s}\right\rangle_{c}$ are partial vacua of the
quasi-photon of the first ($s=1$) and the second kind ($s=2$), respectively.

Let us consider the case where the parameter $\epsilon$ is small, in
particular,
\begin{equation}
	\epsilon\ll\Delta\kappa=\left\vert \kappa_{2}-\kappa_{1}\right\vert\,.
	\label{1.29}
\end{equation}
Then approximate solutions of the characteristic equation (\ref{1.27}) read:
\begin{equation}
	r_{s}=\kappa_{s}+\frac{\epsilon}{2\kappa_{s}}+\frac{2\kappa_{s}+\kappa_{1}
		^{2}+\kappa_{2}^{2}}{8\kappa_{s}^{3}}\sum_{l\neq s}\frac{\epsilon^{2}}
	{\kappa_{s}^{2}-\kappa_{l}^{2}}+O(\epsilon^{3})\,, 
	\nonumber
\end{equation}
such that:
\begin{eqnarray}\nonumber
	u_{s,s^{\prime}}&=&\delta_{s,s^{\prime}}+\frac{(1-\delta_{s,s^{\prime}}
		)\epsilon}{2\sqrt{\kappa_{1}\kappa_{2}}(\kappa_{s^{\prime}}-\kappa_{s}
		)}+O(\epsilon^{2}),\\  
	v_{s,s^{\prime}}&=&\frac{\epsilon}{2\sqrt{\kappa
			_{1}\kappa_{2}}(\kappa_{s^{\prime}}+\kappa_{s})}+O(\epsilon^{2})\,.
	\label{2.2}
\end{eqnarray}

\subsection{The system in a constant uniform magnetic field\label{S2.3}}

Here we consider the system under consideration placed in the external
constant and uniform magnetic field $\mathbf{B}$ directed along the photon
beam, $\mathbf{B}=(0,0,B>0)$. In fact, the external field
affects directly only electrons, but then, due to the electron--photon
interaction, it affects photons as well.

The Hamiltonian of the system has the form (\ref{sc1b}), where
$\mathbf{A_{\mathrm{ext}}(r})$ is the vector potential of the constant
magnetic field taken in the Landau gauge, 
$\mathbf{A}_{\mathrm{ext}}(\mathbf{r})=(-Bx^{2},0,0)$. In this case, equation
(\ref{1.14}) admits three integrals of motion $\hat{G}_{\mu}$, $\mu=0$, $1$,
$3$, where $\hat{G}_{\mu}$ are given by Eqs. (\ref{5.2}). Then we look for
solutions of Eq. (\ref{1.14}) that are eigenvectors for the integrals of
motion,
\begin{equation}
	\hat{G}_{\mu}\Phi_{\mathrm{\gamma}}(x)=g_{\mu}\Phi_{\mathrm{\gamma}
	}(x),\ \ \mu=0,1,3\,. 
	\label{4.1}
\end{equation}
With account taken of Eqs. (\ref{4.1}) one can see that vectors 
$\Phi_{\mathrm{\gamma}}(x)$ satisfy the following equation:
\begin{align}
	&  \left[  \hat{H}_{\mathrm{K}}^{\prime}-\frac{g_{0}-g_{3}}{2}\right]
	\Phi_{\mathrm{\gamma}}(x)=0\ ,\label{4.2}\\
	&  \ \hat{H}_{\mathrm{K}}^{\prime}=\hat{H}_{\mathrm{\gamma}}+\frac{1}
	{2(ng)}\left\{  \left[  eBx^{2}-g^{1}-e\hat{A}^{1}(u)\right]  ^{2}+\left[
	i\partial_{2}-e\hat{A}^{2}(u)\right]  ^{2}+m^{2}\right\}\,. 
	\nonumber
\end{align}
The latter fact implies that the operator $\hat{H}_{\mathrm{K}}'$ commutes with
the operator $\hat{P}_{\mu}\hat{P}^{\mu}-m$ on solutions of Eq. (\ref{4.2}),
therefore, it is also an integral of motion.

Let us suppose that vectors $\Phi_{\mathrm{\gamma}}(x)$ are solutions of the
eigenvalue problem
\begin{equation}
	\hat{H}_{\mathrm{K}}^{\prime}\Phi_{\mathrm{\gamma}}(x)=E_{\mathrm{K}}
	\Phi_{\mathrm{\gamma}}(x)\,. 
	\label{4.4}
\end{equation}
Then Eq. (\ref{4.2}) is satisfies identically, if $g_{0}=E_{\mathrm{K}%
}+ng/2$. It follows from Eq. (\ref{4.1}) that
\begin{equation}
	\Phi_{\mathrm{\gamma}}(x)=\exp\left[-i(g_{0}t+g_{1}x^{1}+g_{3}z)\right]
	\hat{U}\Phi_{\mathrm{K}}(x^{2})\,, 
	\label{4.6}
\end{equation}
where the operator $\hat{U}$ is given by Eq. (\ref{5.9}). Substituting Eq.
(\ref{4.6}) into Eq. (\ref{4.4}), we obtain:
\begin{align}
	&  \hat{H}_{\mathrm{K}}\Phi_{\mathrm{K}}(x^{2})=E_{\mathrm{K}}\Phi
	_{\mathrm{K}}(x^{2})\ ,\nonumber\\
	&  \hat{H}_{\mathrm{K}}=\hat{H}_{\mathrm{\gamma}}+\frac{1}{2(ng)}\left[
	\left(  eBx^{2}-g^{1}-\hat{Q}^{1}\right)  ^{2}+(i\partial_{2}-\hat{Q}^{2}
	)^{2}+m^{2}\right]\,. 
	\label{4.8}
\end{align}

The operator $\hat{G}_{\mu}$ can be represented in the form (\ref{5.5}),
where the operator $\hat{H}_{\mathrm{K}}$ is given by Eq. (\ref{4.8}). We may
find solutions of Eqs. (\ref{1.14}) and (\ref{4.1}), that are at the same time
eigenvectors of the integrals of motion $(\hat{P}_{\mathrm{K}})_{\mu}$ and
$\hat{H}_{\mathrm{K}}$:
\begin{eqnarray}
	  (\hat{P}_{\mathrm{K}})_{\mu}\ \Phi_{\mathrm{\gamma}}(x)&=&p_{\mu}
	\Phi_{\mathrm{\gamma}}(x),\ \ \ \ \mu=0,1,3\ ,\label{4.8a}\\
	  \hat{H}_{\mathrm{K}}\Phi_{\mathrm{\gamma}}(x)&=&E_{\mathrm{ph}}
	\Phi_{\mathrm{\gamma}}(x)\,, 
	\label{4.8b}
\end{eqnarray}
It follows from Eqs. (\ref{4.1}) and (\ref{4.8a})-(\ref{4.8b}) that%
\begin{equation}
	g_{0}=p_{0}+E_{\mathrm{ph}},\ \ g_{1}=p_{1},\ \ g_{3}=p_{3}-E_{\mathrm{ph}%
	},\ \ ng=np\,. 
	\nonumber
\end{equation}

At this stage, we, following the pioneer work by Malkin and Man`ko \cite{154},
introduce new creation $\hat{c}_{0}^{\dagger}$ and annihilation $\hat{c}_{0}$
operators, $[\hat{c}_{0},\hat{c}_{0}^{\dagger}]=1$,
\begin{align}
	&  \hat{c}_{0}=(2)^{-1/2}\left(  \eta+\partial_{\eta}\right)  ,\ \ \hat{c}
	_{0}^{\dagger}=(2)^{-1/2}\left(  \eta-\partial_{\eta}\right)  ,\ (eB)^{1/2}
	\eta=\left(  eBx^{2}-g^{1}\right)  \ ,\nonumber\\
	&  x^{2}=\sqrt{\frac{1}{2eB}}\left(  \hat{c}_{0}-\hat{c}_{0}^{\dagger}\right)
	+g^{1},\ \ \partial_{2}=\sqrt{\frac{eB}{2}}\left(  \hat{c}_{0}+\hat{c}
	_{0}^{\dagger}\right)\,, 
	\nonumber
\end{align}
that, in fact, describe circular motion of charged particles in the magnetic
field. The latter operators commute with all the quasi-photon operators
$\hat{c}_{s\mathbf{,}\lambda}$ and $\hat{c}_{s^{\prime},\lambda^{\prime}}^{\dagger}$, $s,s^{\prime}=1,2$. It is convenient to introduce the operators
$\hat{c}_{0,\lambda}^{\dagger}=\hat{c}_{0}^{\dagger}\delta_{\lambda,1}$ and
$\hat{c}_{0,\lambda}=\hat{c}_{0}\delta_{\lambda,1}$.

One can see that the Hamiltonian $\hat{H}_{\mathrm{K}}$ is a quadratic
operator with respect to the extended set of creation $\hat{c}_{s,\lambda
}^{\dagger}$ and annihilation $\hat{c}_{s,\lambda}$, $s=0,1,2$,$\ \lambda=1,2$
operators. Then with the help of the linear canonical transformation
(\ref{5.13}), one can diagonalize the Hamiltonian $\hat{H}_{\mathrm{K}}$. 
The diagonalized Hamiltonian is represented as a sum of two commuting terms, 
the one $H_{\mathrm{e}}$ corresponds to the electron subsystem, while the second one $H_{\mathrm{ph}}$ to the subsystem of the quasi-photons,
\begin{eqnarray}
	  \hat{H}_{\mathrm{K}}&=&H_{\mathrm{ph}}+H_{\mathrm{e}},\ \ H_{\mathrm{e}}
	=\tau_{0}\hat{c}_{0}^{\dagger}\hat{c}_{0}+\left[  m^{2}(np)^{-1}
	-\omega\right]  /2,\ \omega=eB(np)^{-1}\ ,\ 
	\nonumber\\
	  H_{\mathrm{ph}}&=&\sum_{s=1,2}\sum_{\lambda=1,2}\tau_{s}c_{s,\lambda
	}^{\dagger}c_{s,\lambda}-\sum_{s=1,2}\sum_{\lambda,\lambda^{\prime}=1,2}
	\tau_{k,\lambda^{\prime}}\left\vert v_{s,\lambda;k,\lambda^{\prime}
	}\right\vert ^{2}+\frac{\epsilon}{2}\left(  \kappa_{1}^{-1}+\kappa_{2}
	^{-1}\right)\,.
	\nonumber
\end{eqnarray}
Eigenvalues $E_{\mathrm{K}}$ and $E_{\mathrm{ph}}$ are:
\begin{eqnarray}
	E_{\mathrm{K}}&=&E_{\mathrm{ph}}+\tau_{0}N_{0}\ +\left[  m^{2}(np)^{-1}
	+\omega\right]  /2\,,\nonumber\\  
	E_{\mathrm{ph}}&=&\sum_{s=1,2}\sum_{\lambda=1,2}\tau
	_{s}N_{s,\lambda}+H_{0},\ N_{0}\in\mathbb{N},\ \ N_{s,\lambda}\in
	\mathbb{N}\,.
	\label{4.12c}
\end{eqnarray}

With account taken of Eqs. (\ref{4.12c}), we obtain from the equation
$g_{0}=E_{\mathrm{K}}+np/2$ that
\begin{equation}
	p_{0}^{2}=eB\left(  2\frac{\tau_{0}}{\omega}N_{0}+1\right)+p_{3}^{2}
	+m^{2}\,.
	\nonumber
\end{equation}

Quantities $r_{k,\lambda}$ are positive roots of the equation
\begin{equation}
	\sum_{s=1,2}\frac{\epsilon}{\tau_{k,\lambda}^{2}-\kappa_{s}^{2}}
	=1+\frac{\left(  -1\right)  ^{\lambda-1}\omega}{\tau_{k,\lambda}}
	,\ \ \tau_{0,\lambda}=\tau_{0}\delta_{\lambda,1},\ \ k=0,1,2\,, 
	\label{4.13}
\end{equation}
satisfying the conditions $\tau_{k,\lambda}(\epsilon=0)=\kappa_{s}$.

In what follows, we only need explicit expressions for the matrices
$u_{s,\lambda;k,\lambda^{\prime}}$ and $v_{s,\lambda;k,\lambda^{\prime}}$ with
$s,k=1,2$, and $\lambda =1,2$. They are:
\begin{eqnarray}
	  u_{s,\lambda;k,\lambda^{\prime}}&=&\left[  \left(  \sqrt{\frac{\tau
			_{k,\lambda^{\prime}}}{\kappa_{s}}}+\sqrt{\frac{\kappa_{s}}{\tau
			_{k,\lambda^{\prime}}}}\right)  \frac{(-1)^{\lambda^{\prime}-1}\delta
		_{\lambda,1}-i\delta_{\lambda,2}}{2\left(  \tau_{k,\lambda^{\prime}}
		-\kappa_{s}^{2}\right)  }\right]  q_{k,\lambda^{\prime}}\ ,\label{4.13a}\\
	  v_{s,\lambda;k,\lambda^{\prime}}&=&\left[  \left(  \sqrt{\frac{\tau
			_{k,\lambda^{\prime}}}{\kappa_{s}}}-\sqrt{\frac{\kappa_{s}}{\tau
			_{k,\lambda^{\prime}}}}\right)  \frac{(-1)^{\lambda^{\prime}-1}\delta
		_{\lambda,1}-i\delta_{\lambda,2}}{2\left(  \tau_{k,\lambda^{\prime}}
		-\kappa_{s}^{2}\right)  }\right]  q_{k,\lambda^{\prime}}\ ,\label{4.13b}\\
	  q_{k,\lambda}&=&\left[  \frac{\left(  -1\right)  ^{\lambda}\omega}
	{\tau_{k,\lambda}^{3}\epsilon}+2\sum_{s=1,2}\left(  \tau_{k,\lambda}
	-\kappa_{s}^{2}\right)  ^{-2}\right]  ^{-1/2}\,. 
	\label{4.13c}
\end{eqnarray}

Stationary states of the complete system have the form $\left\vert
\Phi_{\mathrm{ph}}\right\rangle \otimes\left\vert \Phi_{\mathrm{e}}\right\rangle$, 
where $\left\vert \Phi_{\mathrm{ph}}\right\rangle$ are
states (\ref{1.28a}) of the quasi-photons and 
$\left\vert \Phi_{\mathrm{e}}\right\rangle$ are some states vectors of the electron subsystem, explicit forms of which is not important for our purposes.

In what follows, we consider the parameter $\epsilon$ to be small, in
particular, it satisfies Eq. (\ref{1.29}). In this case, it follows from Eq. (\ref{4.13}) that
\begin{equation}
	\tau_{k,\lambda}=
	\kappa_{k}+\frac{\epsilon/2}{\kappa_{k}+(-1)^{\lambda-1}\omega}+O(\epsilon^{2}),\ \ k,\lambda=1,2\,. 
	\nonumber
\end{equation}

\section{Entanglement of the photon beam by the electron medium and external
	magnetic field\label{S3}}

Below, we use exact solutions of the model for calculating entanglement of the
photon beam by quantized electron medium and by a constant magnetic field. In
this respect, we recall that the entanglement is a genuine quantum property which
is associated with a quantum non-separability of parts of a composite system.
Entangled states became a powerful tool for studying principal questions both
in quantum theory and in quantum computation and information theory
\cite{Bell,Preskill,NieCh00,BellB,OliBoS}. Quantum entanglement has
applications in emerging quantum computing and quantum cryptography
technologies, and has been used to perform quantum teleportation
experimentally. We note that peremptory experimental confirmation of the
existence of the quantum entanglement is presented e.g. in the latter
references. In particular, it was studied a measuring correlations between
states of electronic spins in diamonds, in which the violation of a Bell
inequality was verified. Recently it was proposed a two-qubit photonic quantum
processor that implements two consecutive quantum gates on the same pair of
polarization-encoded qubits \cite{Barz}. Different views on what is actually
happening in the process of quantum entanglement may be related to different
interpretations of quantum mechanics. We believe that the complete
understanding of the nature of quantum entanglement still requires a detailed
consideration of a variety of relatively simple cases, not only in
nonrelativistic quantum mechanics, but in QFT as well. This explains recent
interest in study general problems of quantum entanglement in QFT
\cite{Nishi18,Witte18}, and in considering specific examples in QFT of systems
with unstable vacuum
\cite{CamPa05,LinChH10,AdaBuP13,BusP13,BruFrF13,FinCa13,302}. Some
possibilities to control the degree of entanglement of photon beams by
changing external conditions were considered in Refs. \cite{294,323,329}.

\subsection{Entanglement in the absence of an external field\label{S3.1}}

Here we calculate the entanglement of the photon beam only by the electron
medium in the absence of an external field, using results presented in Sec.
\ref{S2.2}. We recall, that in this case transversal momenta of all the
electrons from the electron medium are chosen to be zero.

As a result of comparatively cumbersome calculations, it can be seen that
\begin{equation}
	\hat{c}_{1,\lambda_{1}}^{+}\hat{c}_{2,\lambda_{2}}^{+}\left\vert
	0\right\rangle _{c}=\hat{c}_{1,\lambda_{1}}^{+}\hat{c}_{2,\lambda_{2}}
	^{+}\left\vert 0\right\rangle +M ,\ \  M =O\left(  \epsilon\right)
	\,, 
	\label{3.1}
\end{equation}
where $M$ is a vector of many (more than two) photon states, and 
$\left\vert 0\right\rangle _{c}=\left\vert 0_{1}\right\rangle_{c}\otimes\left\vert
0_{2}\right\rangle _{c}$ is the vacuum of the quasi-photons.

We believe that a free photon nonentangled beam after passing through a macro
region filled with charged particles, is deformed namely to the form
(\ref{3.1}). The vector 
$\hat{c}_{1,2}^{+}\hat{c}_{2,1}^{+}\left\vert 0\right\rangle$ represents a two-photon entangled state of initial free photons. Being normalized this vector reads:
\begin{equation}
	C \hat{c}_{1,\lambda_{1}}^{+}\hat{c}_{2,\lambda_{2}}^{+}\left\vert
	0\right\rangle =\left\vert \lambda_{1},\lambda_{2}\right\rangle
	,\ C=\left\langle 0\right\vert \hat{c}_{2,\lambda_{2}}\hat{c}_{1,\lambda_{1}
	}\hat{c}_{1,\lambda_{1}}^{+}\hat{c}_{2,\lambda_{2}}^{+}\left\vert
	0\right\rangle ^{-1/2}\,, 
	\nonumber
\end{equation}
where $C$ is a normalization constant. Let's assume that it is possible to
have an analyzer extracting this state for measuring its entanglement. Then
the measure of the entanglement of the initial photon beam can be identify
with the measure of the entanglement of the state 
$\left\vert \lambda_{1},\lambda_{2}\right\rangle $. 
Below we calculate von Neumann and Schmidt measure (see Appendix \ref{S5}) 
of such an entanglement.

With account taken of Eqs. (\ref{5.13})-(\ref{1.25}), we obtain:
\begin{equation}
	\left\vert \lambda_{1},\lambda_{2}\right\rangle =C \sum_{s,s^{\prime}
	}u_{s^{\prime},1}u_{s,2}\hat{a}_{s,\lambda_{1}}^{\dagger}\hat{a}_{s^{\prime
		},\lambda_{2}}^{\dagger}\,,
	\label{3.2c}
\end{equation}
It follows from Eq. (\ref{2.2}) that for $\Delta^{-1}\kappa\ll1$ contributions
of the terms $u_{1,1}u_{1,2}\hat{a}_{1,\lambda_{1}}^{\dagger}\hat
{a}_{1,\lambda_{2}}^{\dagger}\left\vert 0\right\rangle $ and 
$u_{2,1}u_{2,2}\hat{a}_{2,\lambda_{1}}^{\dagger}\hat{a}_{2,\lambda_{2}}^{\dagger
}\left\vert 0\right\rangle$ to state (\ref{3.2c}) are small compared to other
contributions. Then
\begin{align}
	&  \left\vert \lambda_{1},\lambda_{2}\right\rangle =C \left(
	u_{2,1}u_{1,2}\hat{a}_{1,\lambda_{2}}^{\dagger}\hat{a}_{2,\lambda_{1}
	}^{\dagger}+u_{1,1}u_{2,2}\hat{a}_{2,\lambda_{2}}^{\dagger}\hat{a}
	_{1,\lambda_{1}}^{\dagger}\right)  \left\vert 0\right\rangle \ +O\left(
	\Delta^{-1}\kappa\right)\,,
	\nonumber\\
	&  C =\left[  \left(  u_{2,1}u_{1,2}\right)  ^{2}+\left(  u_{1,1}%
	u_{2,2}\right)  ^{2}\right]  ^{-1/2}\,.
	\nonumber
\end{align}

In the case under consideration, the computational bases (\ref{a.2}) can be
chosen as:
\begin{equation}
	\left\vert \theta_{1}\right\rangle =\hat{a}_{1,\lambda_{2}}^{\dagger}\hat
	{a}_{2,\lambda_{1}}^{\dagger}\left\vert 0\right\rangle ,\ \ \left\vert
	\theta_{2}\right\rangle =\hat{a}_{2,\lambda_{2}}^{\dagger}
	\hat{a}_{1,\lambda_{1}}^{\dagger}\left\vert 0\right\rangle \,.
	\nonumber
\end{equation}
In terms of such a bases the states $\left\vert \lambda_{1},\lambda_{2}\right\rangle$ are:
\begin{align}
	&  \left\vert \lambda_{1},\lambda_{2}\right\rangle =\sum_{j=1}^{2}
	v_{j}\left\vert \theta_{j}\right\rangle \ ,\nonumber\\
	&  v_{1}=\frac{u_{2,1}u_{1,2}}{\sqrt{\left(  u_{2,1}u_{1,2}\right)
			^{2}+\left(  u_{1,1}u_{2,2}\right)  ^{2}}},\ v_{2}=\frac{u_{1,1}u_{2,2}}
	{\sqrt{\left(  u_{2,1}u_{1,2}\right)  ^{2}+\left(  u_{1,1}u_{2,2}\right)
			^{2}}}\,. 
	\nonumber
\end{align}

The density matrix $\hat{\rho}$, corresponding to the pure state 
$\left\vert\lambda_{1},\lambda_{2}\right\rangle $ has the form 
$\hat{\rho}=\left\vert\lambda_{1},\lambda_{2}\right\rangle \left\langle \lambda_{1},
\lambda_{2}\right\vert$.

The entanglement measure between the subsystems of free photons can be
calculated as the information or Schmidt measure of a reduced density matrix
$\hat{\rho}^{(1)}$ corresponding to the subsystem of the free photon of
first type (the same result can be obtained by using the reduced density
matrix $\hat{\rho}^{(2)}$, corresponding to the subsystem of the free
photon of second type). The reduced density matrix $\hat{\rho}^{(1)}$ can
be obtained by calculating the trace of the general density matrix $\hat{\rho}$ 
over the subsystem of free photons of the second type,
\begin{equation}
	\hat{\rho}^{(1)}=\mathrm{tr}_{2}\hat{\rho}=\sum_{\lambda=1,2}
	\left\langle 0_{1}\right\vert \hat{a}_{2,\lambda} \hat{\rho} 
	\hat {a}_{2,\lambda}^{\dagger} \left\vert 0_{1}\right\rangle\,, 
	\label{3.10}
\end{equation}
where $\mathrm{tr}_{s}\hat{\rho}$ is partial trace has been taken over one
subsystem, either first type $s=1$ or $s=2$. By $\mathrm{tr}\hat{\rho}$
without a subscript, the complete trace is denoted.

For two photons with the same polarizations, $\lambda_{1}=\lambda_{2}$ the
reduced density matrix $\hat{\rho}^{(1)}$ represents a pure state,
\begin{equation}
	\left.  \hat{\rho}^{(1)}\right\vert _{\lambda_{1}=\lambda_{2}=\lambda
	}=\left(  v_{1}^{2}+v_{2}^{2}\right)  \ \left\vert 1,\lambda\right\rangle
	\left\langle 1,\lambda\right\vert ,\ \ \left\vert 1,\lambda\right\rangle
	=\hat{a}_{1,\lambda}^{\dagger}\ \left\vert 0_{1}\right\rangle \,. 
	\nonumber
\end{equation}
For two photons with different polarizations, we have an entangled state
\begin{equation}
	\left.  \hat{\rho}^{(1)}\right\vert _{\lambda_{1}\neq\lambda_{2}}=
	v_{1}^{2}\ \left\vert 1,\lambda_{2}\right\rangle \left\langle 1,\lambda
	_{2}\right\vert +v_{2}^{2}\ \left\vert 1,\lambda_{1}\right\rangle \left\langle
	1,\lambda_{1}\right\vert ,\ \ \mathrm{tr}\hat{\rho}^{(1)}=1\,. 
	\label{3.12}
\end{equation}

Representation (\ref{3.12}) for the reduced density matrix $\hat{\rho}^{(1)}$ 
allows us to calculate the corresponding von Neumann entropy (\ref{a.3}),
\begin{equation}
	S\left(  \hat{\rho}^{(1)}\right)  =-\mathrm{tr}\left(  \hat{\rho}
	^{(1)}\ln\hat{\rho}^{(1)}\right)  =-\sum_{a=1,2}\beta_{a}\log_{2}
	\ \beta_{a}\,, 
	\nonumber
\end{equation}
where $\beta_{a}$, $a=1$, $2$, are eigenvalues of the operator $\hat{\rho}^{(1)}$,
\begin{align}
	&  \beta_{2}=1-\beta_{1}\ ,\ \ \beta_{1}=\frac{\left(  u_{2,1}u_{1,2}\right)
		^{2}}{\left(  u_{2,1}u_{1,2}\right)  ^{2}+\left(  u_{1,1}u_{2,2}\right)  ^{2}
	}=\left\{  1+\right. \nonumber\\
	&  \left.  \left[  \frac{\kappa_{1}-\kappa_{2}+\mathrm{sgn}(\kappa_{1}
		-\kappa_{2})\sqrt{4\epsilon^{2}+(\kappa_{1}^{2}-\kappa_{2}^{2})^{2}}}
	{\kappa_{1}-\kappa_{2}-\mathrm{sgn}(\kappa_{1}-\kappa_{2})\sqrt{4\epsilon
			^{2}+(\kappa_{1}^{2}-\kappa_{2}^{2})^{2}}}\frac{(r_{1}+\kappa_{1}
		)(r_{2}+\kappa_{2})}{(r_{1}+\kappa_{2})(r_{2}+\kappa_{1})}\right]
	^{2}\right\}  ^{-1}=\nonumber\\
	& \frac{\left(  \epsilon/\Delta\kappa\right)  ^{4}}
	{(4\kappa_{1}\kappa_{2})^{2}}+O(\epsilon^{5})\,, 
	\label{3.14a}
\end{align}
The asymptotic behavior of the information measure follows from Eq. (\ref{3.14a}),
\begin{align}
	& S\left(  \hat{\rho}^{(1)}\right)  =\frac{\Phi_{0}\left(  2,1\right)  }
	{2\ln2}\left[  \epsilon^{4}\left(  1-\ln\left(  \Phi_{0}\left(  2,1\right)
	/2\right)  \right)  -4\epsilon^{4}\ln\epsilon\right],\nonumber\\
	& \Phi_{0}\left(2,1\right)  =\frac{2(\Delta\kappa)^{-4}}{(4\kappa_{1}\kappa_{2})^{2}}\,.
	\nonumber
\end{align}

In the case under consideration the Schmidt measure (\ref{s.10}) reads:
\begin{equation}
	E_{\mathrm{S}}(\Psi(\lambda_{1},\lambda_{2}))=\epsilon^{4}\Phi_{0}
	\left(2,1\right)\,.
	\nonumber
\end{equation}

\subsection{Entanglement in the presence of a constant uniform magnetic
	field\label{S3.2}}

Here we calculate the entanglement of the photon beam by the electron medium
in the presence of the external constant magnetic field, using results
presented in Sec. \ref{S2.3}.

Consider the state vector $\left\vert \Phi_{\mathrm{ph}}\right\rangle$ with
only two quasi-photons, one of the first kind, and another of the second kind,
and with anti-parallel polarizations, which we take as $\lambda_{1}=2$
and\ $\lambda_{2}=1$. Such a state vector corresponds to $N_{1,1}=N_{2,2}=0$
and $N_{1,2}=N_{2,1}=1$ and has the form 
$\hat{c}_{1,2}^{\dagger}\hat{c}_{2,1}^{\dagger}\left\vert 0\right\rangle _{c}$. 
As a result of comparatively cumbersome calculations, it can be seen that in this case
\begin{equation}
	\hat{c}_{1,2}^{\dagger}\hat{c}_{2,1}^{\dagger}\left\vert 0\right\rangle
	_{c}=\hat{c}_{1,2}^{\dagger}\hat{c}_{2,1}^{\dagger}\left\vert 0\right\rangle
	+M,\ \  M =O\left(  \epsilon^{2}\right)\,, 
	\label{3.21}
\end{equation}
where $M$ is a vector of many (more than two) photon states, and 
$\left\vert 0\right\rangle _{c}=\left\vert 0_{1}\right\rangle_{c}\otimes\left\vert
0_{2}\right\rangle _{c}$ is the vacuum of the quasi-photons. We believe that a
free photon nonentangled beam after passing through a macro region filled with
charged particles moving in the magnetic field, is deformed namely to the form
(\ref{3.21}). The vector $\hat{c}_{1,2}^{+}\hat{c}_{2,1}^{+}\left\vert
0\right\rangle$ represents a two-photon entangled state of initial free photons.
\begin{equation}
	\left\vert 2,1\right\rangle =C \ \hat{c}_{1,2}^{\dagger}\hat{c}
	_{2,1}^{\dagger}\left\vert 0\right\rangle ,\ \ C=\left\langle 0\right\vert
	\hat{c}_{2,1}\hat{c}_{1,2}\hat{c}_{1,2}^{\dagger}\hat{c}_{2,1}^{\dagger
	}\left\vert 0\right\rangle ^{-1/2}\,. 
	\label{3.22}
\end{equation}
Let's assume that it is possible to have an analyzer extracting this state for
measuring its entanglement. Then the measure of the entanglement of the
initial photon beam can be identify with the measure of the entanglement of
the state (\ref{3.22}). Below we calculate von Neumann and Schmidt measure
(see Appendix \ref{S5}) of such an entanglement.

With account taken of Eqs. (\ref{5.13}), (\ref{4.13a})-(\ref{4.13c}), we
obtain:
\begin{align}
	  & \left\vert 2,1\right\rangle =
	  C\sum_{s=1,2}\sum_{\lambda=1,2}u_{s,\lambda}\tilde{u}_{s^{\prime},\lambda^{\prime}}\hat{a}_{s,\lambda}^{\dagger}\hat{a}_{s^{\prime},\lambda^{\prime}}^{\dagger}\left\vert 0\right\rangle,\nonumber\\ 
	  & u_{s,\lambda}=u_{s,\lambda;1,2},\quad 
	  \tilde{u}_{s,\lambda}=u_{s,\lambda;2,1}\,. 
	  \label{f17}
\end{align}
One can see that in the approximation under consideration we have to neglect
terms of the form $u_{1,\lambda}\tilde{u}_{1,\lambda^{\prime}}a_{1,\lambda
}^{+}a_{1,\lambda^{\prime}}^{+}\left\vert 0\right\rangle $ and $u_{2,\lambda
}\tilde{u}_{2,\lambda^{\prime}}a_{2,\lambda}^{+}a_{2,\lambda^{\prime}}
^{+}\left\vert 0\right\rangle $ (that correspond to transitions with the same
frequencies) in the right hand side of Eq. (\ref{f17}). Thus, we obtain
\begin{align}
	&  \left\vert 2,1\right\rangle =C\sum_{\lambda,\lambda^{\prime}}\left[
	u_{1,\lambda}\tilde{u}_{2,\lambda^{\prime}}+u_{2,\lambda^{\prime}}\tilde
	{u}_{1,\lambda}\right]  \hat{a}_{1,\lambda}^{+}\hat{a}_{2,\lambda^{\prime}
	}^{+}\left\vert 0\right\rangle +O\left(  \left(  \Delta\kappa\right)
	^{-1}\right)  \ ,\label{f17a}\\
	&  C=\left(  \sum_{\lambda,\lambda^{\prime}}\left\vert u_{1,\lambda}\tilde
	{u}_{2,\lambda^{\prime}}+u_{2,\lambda^{\prime}}\tilde{u}_{1,\lambda
	}\right\vert ^{2}\right)  ^{-1/2}\,.
	\nonumber
\end{align}
Introducing the computational basis,
\begin{align}
	& \left\vert \vartheta_{1}\right\rangle =a_{1,1}^{+}a_{2,1}^{+}\left\vert
	0\right\rangle,\quad \left\vert \vartheta_{2}\right\rangle =a_{1,1}^{+}
	a_{2,2}^{+}\left\vert 0\right\rangle ,\nonumber\\ 
	& \left\vert \vartheta_{3}%
	\right\rangle =a_{1,2}^{+}a_{2,1}^{+}\left\vert 0\right\rangle ,\ \,\, \ \left\vert
	\vartheta_{4}\right\rangle =a_{1,1}^{+}a_{2,1}^{+}\left\vert 0\right\rangle\,, 
	\nonumber
\end{align}
we rewrite vector (\ref{f17a}) as:
\begin{align}
	& \left\vert 2,1\right\rangle = 
	  C\sum_{j=1}^{4}\upsilon_{j}\left\vert\vartheta_{j}\right\rangle,\quad
	  C=\left(\sum_{i=1}^{4} \mid \upsilon_{i} \mid^{2} \right)^{-1/2},\nonumber\\
	&  \upsilon_{1}=u_{1,1}\tilde{u}_{2,1}+u_{2,1}\tilde{u}_{1,1},\ \ \upsilon
	_{2}=u_{1,1}\tilde{u}_{2,2}+u_{2,2}\tilde{u}_{1,1}\ ,\nonumber\\
	&  \upsilon_{3}=u_{1,2}\tilde{u}_{2,1}+u_{2,1}\tilde{u}_{1,2},\ \ \upsilon
	_{4}=u_{1,2}\tilde{u}_{2,2}+u_{2,2}\tilde{u}_{1,2}\,. 
	\nonumber
\end{align}
We use the explicit forms of the matrices $u$ from Eq. (\ref{4.13a}) and of
the square roots $r_{k\lambda}$ from Eq. (\ref{4.13}) to calculate the
quantities $\upsilon_{i}$. They latter are:
\begin{align}
	& \upsilon_{1}=-(a+b),\quad \upsilon_{2}=-i(a-b),\quad \upsilon_{3}=-\upsilon
	_{2},\nonumber \\
	&\upsilon_{4}=\upsilon_{1}\ ,\ \ C=\frac{1}{2}\left\vert a^{2}
	+b^{2}\right\vert ^{-1}\, 
	\nonumber
\end{align}
where
{\footnotesize
\begin{align}
	&  a=\frac{\left(  \sqrt{\frac{\kappa_{1}}{\tau_{21}}}+\sqrt{\frac{\tau_{21}
			}{\kappa_{1}}}\right)  \left(  \sqrt{\frac{\kappa_{2}}{\tau_{12}}}+\sqrt
		{\frac{\tau_{12}}{\kappa_{2}}}\right)  }{4\left(  \tau_{12}^{2}-\kappa_{2}
		^{2}\right)  \left(  \tau_{21}^{2}-\kappa_{1}^{2}\right)  \sqrt{\frac
			{2}{\left(  \tau_{21}^{2}-\kappa_{1}^{2}\right)  ^{2}}+\frac{2}{\left(
				\tau_{21}^{2}-\kappa_{2}^{2}\right)  ^{2}}-\frac{\omega}{\tau_{21}^{3}
				\epsilon}}\sqrt{\frac{2}{\left(  \tau_{12}^{2}-\kappa_{1}^{2}\right)  ^{2}
			}+\frac{2}{\left(  \tau_{12}^{2}-\kappa_{2}^{2}\right)  ^{2}}+\frac{\omega
			}{\tau_{12}^{3}\epsilon}}}\ ,\nonumber\\
	&  b=\frac{\left(  \sqrt{\frac{\kappa_{1}}{\tau_{12}}}+\sqrt{\frac{\tau_{12}
			}{\kappa_{1}}}\right)  \left(  \sqrt{\frac{\kappa_{2}}{\tau_{21}}}+\sqrt
		{\frac{\tau_{21}}{\kappa_{2}}}\right)  }{4\left(  \tau_{12}^{2}-\kappa_{1}
		^{2}\right)  \left(  \tau_{21}^{2}-\kappa_{2}^{2}\right)  \sqrt{\frac
			{2}{\left(  \tau_{21}^{2}-\kappa_{1}^{2}\right)  ^{2}}+\frac{2}{\left(
				\tau_{21}^{2}-\kappa_{2}^{2}\right)  ^{2}}-\frac{\omega}{\tau_{21}^{3}
				\epsilon}}\sqrt{\frac{2}{\left(  \tau_{12}^{2}-\kappa_{1}^{2}\right)  ^{2}
			}+\frac{2}{\left(  \tau_{12}^{2}-\kappa_{2}^{2}\right)  ^{2}}+\frac{\omega
			}{\tau_{12}^{3}\epsilon}}}\,. 
	\nonumber
\end{align}
}

Let us calculate the information entanglement measure of the state 
$\left\vert 2,1\right\rangle $ using definition (\ref{a.3}) as the von Neumann entropy of
the reduced density operator $\hat{\rho}^{(1)}$ (given by Eq. (\ref{3.10})) of the subsystem of the first photon,
\begin{equation}
	E(\left\vert 2,1\right\rangle )=-\mathrm{tr}\left(  \hat{\rho}^{\left(
		1\right)  }\log_{2}\hat{\rho}^{\left(  1\right)  }\right)  =-\sum
	_{a=1,2}\lambda_{a}\log_{2}\lambda_{a}\,, 
	\nonumber
\end{equation}
where $\lambda_{a}$, $a=1,2$, are eigenvalues of 
$\hat{\rho}^{(1)}$. Then we obtain the quantity $y$ defined by Eq. (\ref{s.8}):
\begin{eqnarray}
	  y&=&\left\vert \frac{a^{2}-b^{2}}{a^{2}+b^{2}}\right\vert =\left\vert
	1-\frac{\epsilon^{4}}{8(\kappa_{1}-\kappa_{2})^{4}(\omega-\kappa_{1}
		)^{2}(\omega+\kappa_{2})^{2}}\right\vert +O(\epsilon^{4})\nonumber\\
	  &=&1-\epsilon^{4}\Phi_{\omega}\left(  2,1\right)  +O(\epsilon^{4}
	),\ \ 0\leq\epsilon^{4}\Phi\left(  2,1\right)  <1\ ,\nonumber\\
	  \Phi_{\omega}\left(  2,1\right)  &=&\frac{\left(  \Delta\kappa\right)  ^{-4}
	}{8(\omega-\kappa_{1})^{2}(\omega+\kappa_{2})^{2}},\ \ \lim_{B\rightarrow
		0}\Phi_{\omega}\left(  2,1\right)  =\Phi_{0}\left(  2,1\right)\,.
	\nonumber
\end{eqnarray}

The asymptotic behavior of the information measure 
$E(\left\vert 1,2\right\rangle )$ as $\epsilon\rightarrow0$ has the form:
\begin{equation}
	E(\left\vert 2,1\right\rangle )=\frac{\Phi_{\omega}\left(  2,1\right)  }
	{2\ln2}\left[  \epsilon^{4}\left(  1-\ln\left(  \Phi_{\omega}\left(
	2,1\right)  /2\right)  \right)  -4\epsilon^{4}\ln\epsilon\right]\,.
	\nonumber
\end{equation}

\begin{figure}[h]
	\centering
	\includegraphics[width=0.6\textwidth]{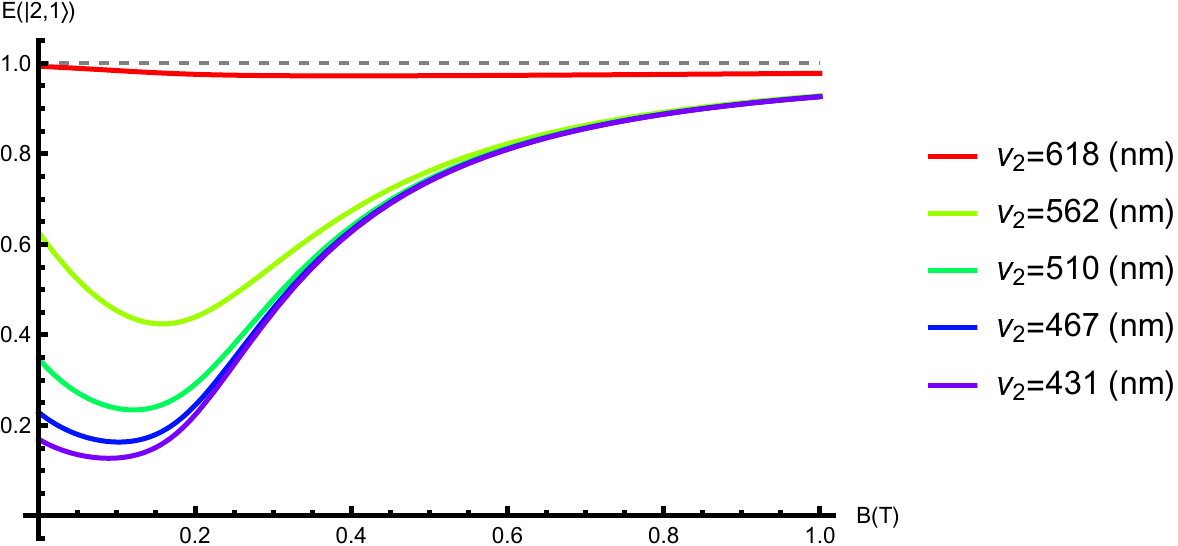}
	\caption{Dependence of the information measure on the parameter $\omega$.}
	\label{graph1}
\end{figure}

Fig 1. shows the behavior of the measure $E(\left\vert 2,1\right\rangle )$
versus the parameter $\omega$ for the first photon of the wavelength $\nu
_{1}=2\pi\kappa_{1}^{-1}=625\ \mathrm{nm}$ and the wavelengths 
$\nu_{2}=2\pi\kappa_{2}^{-1}$ of the second photon. The magnetic field strength $B$
changes from zero to $1\mathrm{T}$, the momentum projection $np=p_{0}-p_{z}$
is chosen as $2.5\cdot10^{7\ }\mathrm{\ m}^{-1}$, the electron density is
chosen as $\rho=2.6\cdot10^{20}\mathrm{el\ m}^{-3}$. Note that the
entanglement decreases with increasing $\Delta\kappa$ for all the values of
$\omega$. We see that turning on of a "weak" magnetic field $B\lesssim
0.1\mathrm{T}$ leads to a decrease of the entanglement, and a further increase
of the magnetic field strength leads to the increase of the entanglement.

Fig 2. shows the behavior of the measure $E(\left\vert 2,1\right\rangle)$ on
the electron density calculated for the wave length of the first photon
$\nu_{1}=2\pi\kappa_{1}^{-1}=625\ \mathrm{nm}$ and for the wave length of the
second photon $\nu_{2}=2\pi\kappa_{2}^{-1}=562\ \mathrm{nm}$, whereas the
parameter $np$ is equal $2.5\cdot10^{7\ }\mathrm{\ m}^{-1}$. We see that an
increase in density leads to an increase in entanglement. But as the magnetic
field increases, the effect of density on entanglement becomes weaker.

\begin{figure}[h]
	\centering
	\includegraphics[width=0.6\textwidth]{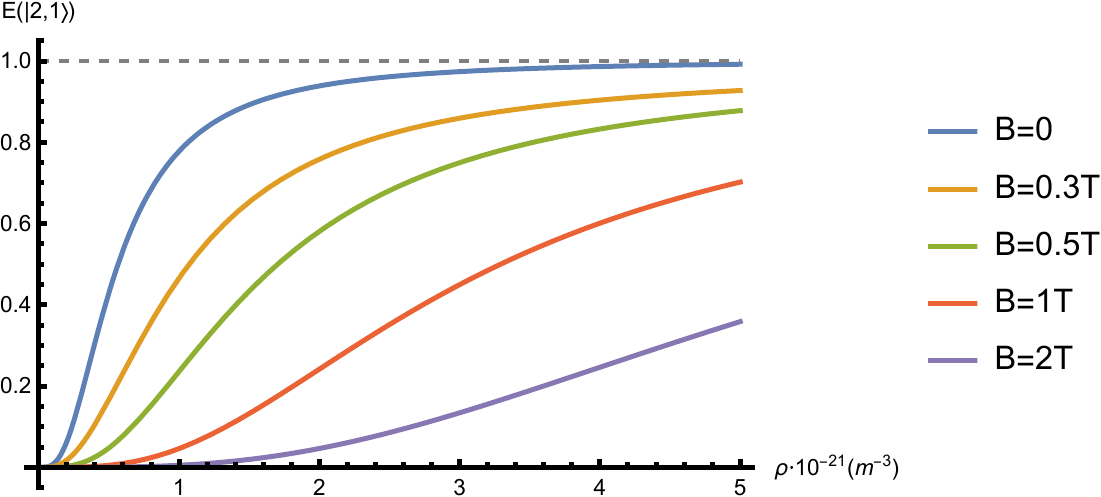}
	\caption{Dependence of the information measure on the electron density
			for different strengths of the magnetic field.}
	\label{graph2}
\end{figure}

An increase in the parameter $np$ leads to a sharp decrease in the effect of
photon entanglement (see. Fig. 3 for $\nu_{1}=2\pi\kappa_{1}^{-1}
=625\ \mathrm{nm}$ and $\nu_{2}=2\pi\kappa_{2}^{-1}=562\ \mathrm{nm}$).

\begin{figure}[h]
	\centering
	\includegraphics[width=0.6\textwidth]{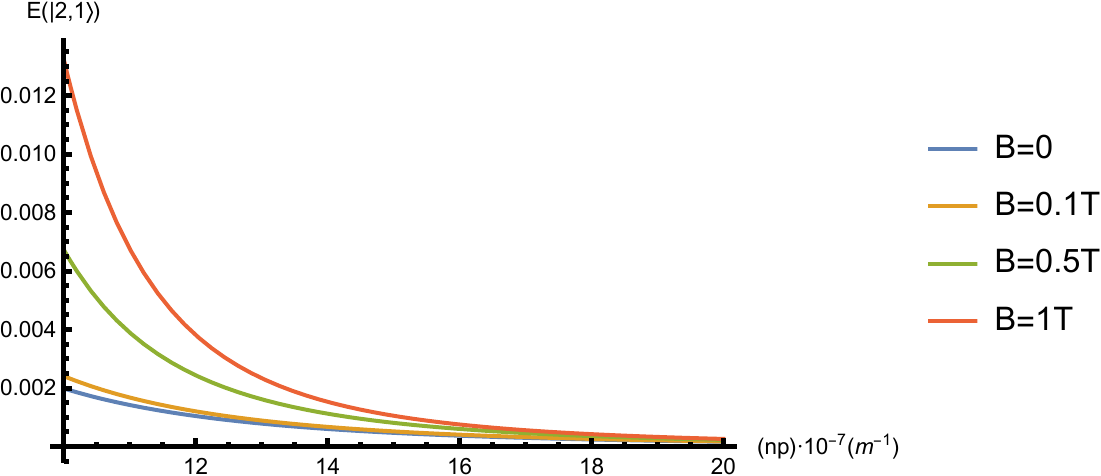}
	\caption{Dependence of the information measure on the parameter $np$.}
	\label{graph3}
\end{figure}

In the case under consideration the Schmidt measure (\ref{s.10}) reads:
\begin{equation}
	E_{\mathrm{S}}(\left\vert 2,1\right\rangle )=\frac{1-y^{2}}{2}\sim\epsilon
	^{4}\Phi\left(  2,1\right)  ,\ \ \epsilon\rightarrow0\,. 
	\nonumber
\end{equation}
Numerical calculations for the the Schmidt measure are given almost by the
same plot as for the information measure.

Let us consider a state $\left\vert \Phi_{\mathrm{ph}}\right\rangle $ with two
quasi-photons, one of the first kind, and another one of the second kind and
with parallel polarizations $\lambda_{1}=1$ and$\ \lambda_{2}=1$. Such a state
vector corresponds to $N_{1,1}=N_{2,1}=1$, $N_{1,2}=N_{2,2}=0$ and has the
form
\begin{equation}
	\left\vert 1,1\right\rangle =C\ \hat{c}_{1,1}^{\dagger}\hat{c}_{2,1}^{\dagger
	}\left\vert 0\right\rangle \ ,\ \ C=\left\langle 0\right\vert \hat{c}
	_{2,1}\hat{c}_{1,1}\hat{c}_{1,1}^{\dagger}\hat{c}_{2,1}^{\dagger}\left\vert
	0\right\rangle ^{-1/2}\,. 
	\nonumber
\end{equation}
One can easily see that in this case, both the information measure of the
state $\left\vert 1,1\right\rangle $ are zero $E(\left\vert 1,1\right\rangle
)=E_{\mathrm{S}}(\left\vert 1,1\right\rangle )=0$. The same result holds true
for the state $\left\vert 2,2\right\rangle =c_{1,2}^{+}c_{2,2}^{+}\left\vert
0\right\rangle $ with two quasi-photons, one of the first kind, and another
one of the second kind and with the parallel polarizations $\lambda_{1}
=2$,$\ \lambda_{2}=2$, $E(\left\vert 2,2\right\rangle )=E_{\mathrm{S}}
(\left\vert 2,2\right\rangle)=0$.

\section{Some concluding remarks\label{S4}}

A model for describing the QED system consisting of a photon beam interacting
with quantized charged spinless media is proposed. The model allows exact
analytical analysis which demonstrates that the above system can be reduced to
two separate subsystems one related to the charged medium and another one
consists of noninteracting quasi-photons. Two versions of the model are
considered one without any external field and another one with a constant
magnetic field. The charged medium is represented by identical relativistic
Bose particles and is characterized by its density and by the energy-momentum
of the particles. We believe that the exact solutions of the model could be
useful in model descriptions such real physical phenomena as photon
entanglement, splitting and fusion of photons (see Ref. \cite{smirnov1974})
and so on. Using the exact solutions of the model, the problem of photon
entanglement is considered. Namely, it is demonstrated that two photons moving
in the same direction with different frequencies and with any of two possible
linear polarizations, can be in a controlled way entangled by passing through
an electron medium without and with applying an external constant magnetic
field. We succeeded to express the corresponding entanglement measures (the
information and the Schmidt ones) via the parameters characteristic of the
problem, such as photon frequencies, magnitude of the magnetic field, and
parameters of the electron medium. We have found that, in the general case,
the entanglement measures depends on the magnitude of the applied magnetic
field and hence can be controlled by the latter. As a rule, the entanglement
increases with increasing the magnetic field (with increasing the cyclotron
frequency). It should be noted that we did not consider resonance cases where
cyclotron frequency approaches photon frequencies. Obviously, the entanglement
depends on the parameters that specify the electron medium such as the
electron density and electron energy and momentum. We did not study this
dependence in detail, these characteristics were fixed by choosing a natural,
small parameter in our calculations. The aim of the latter consideration was
to demonstrate a possibility for entangling photon
beams by the help of an external magnetic field. In contrast with the
well-known possibilities of doing this by using crystal devices, the present
way allows one to change easily and continuously the entanglement measure.

\section{Acknowledgments}

The work is supported by Russian Science Foundation, grant No. 19-12-00042.

\section{Appendix. Entanglement in two-qubit systems}\label{S5}

We recall that a qubit is a two-level quantum-mechanical system with the
Hilbert space $\mathcal{H}=\mathbb{C}^{2}$. Vectors in the space $\mathcal{H}$
are two columns $\left\vert \psi\right\rangle =(\psi_{1},\psi_{2})^{T}$. The scalar product reads: $\langle\psi^{\prime}\left\vert\psi\right\rangle =\psi_{1}^{\prime\ast}\psi_{1}+\psi_{2}^{\prime\ast}\psi_{2}$. 
An orthogonal basis $\left\vert a\right\rangle$, $a=0$, $1$ in
$\mathcal{H}$ can be chosen as: $\left\vert 0\right\rangle=(1,0)^{T}$, 
$\left\vert 1\right\rangle =(0,1)^{T}$,
$\langle a\left\vert a^{\prime}\right\rangle =\delta_{aa^{\prime}}$, 
$\sum_{a=0,1}\left\vert a\right\rangle \langle a\mid\ = I$, 
where $I$ is $2\times2$ unit matrix.

Examples include the spin of the electron in which the two levels can be taken
as spin up and spin down; or the polarization of a single photon in which the
two states can be taken to be the vertical polarization and the horizontal polarization.

Let us consider a system, composed of two qubit subsystems $A$ and $B$ with
the Hilbert space $\mathcal{H}_{AB}=\mathcal{H}_{A}\otimes\mathcal{H}_{B}$
where $\mathcal{H}_{A}=\mathbb{C}^{2}\ $and $\mathcal{H}_{B}=\mathbb{C}^{2}$.
The composite system is a four level and is also called the two-qubit system.
If $\left\vert a\right\rangle_{A}$ and $\left\vert b\right\rangle _{B}$, $a$, 
$b=0$, $1$, are orthonormal bases in the spaces $\mathcal{H}_{A}$ and
$\mathcal{H}_{B}$ respectively, then 
$\left\vert \alpha b\right\rangle=\left\vert a\right\rangle \otimes\left\vert b\right\rangle$ 
is a complete and orthonormalized bases in $\mathcal{H}_{AB}$, which is called
computational bases,
\begin{align}
	&  \ \left\vert \alpha b\right\rangle =\left\vert a\right\rangle
	\otimes\left\vert b\right\rangle =\left(  a_{1}b_{1},a_{1}b_{2},a_{2}
	b_{1},a_{2}b_{2}\right)^{T}\,,
	\nonumber\\
	&  \ \langle\alpha b\left\vert \alpha^{\prime}b^{\prime}\right\rangle
	=\delta_{\alpha\alpha^{\prime}}\delta_{bb^{\prime}}\ ,\ \sum_{\alpha
		,b=0,1}\left\vert \alpha b\right\rangle \langle\alpha b\mid\ =\mathbf{I}\,.
	\nonumber
\end{align}
where $\mathbf{I}$ is $4\times4$ unit matrix. The vectors of the computational
bases are eigenstates of the operator $\sigma_{z}\otimes\sigma_{z}=\mathrm{diag}(1,-1,-1,1)$, 
namely $\left[\sigma_{z}\otimes\sigma_{z}\right]  \left\vert aa^{\prime}\right\rangle =
(-1)^{(a+a^{\prime})}\left\vert aa^{\prime}\right\rangle$. 
The basis is often denoted as $\left\vert \Theta\right\rangle_{s}
$,$\ s=1,2,3,4$,
\begin{align}
	\left\vert \Theta\right\rangle _{1}  &  =\left\vert 00\right\rangle =\left(
	\begin{array}
		[c]{cccc}%
		1 & 0 & 0 & 0
	\end{array}
	\right)  ^{T},\ \ \left\vert \Theta\right\rangle _{2}=\left\vert
	01\right\rangle =\left(
	\begin{array}
		[c]{cccc}%
		0 & 1 & 0 & 0
	\end{array}
	\right)  ^{T}\ ,\nonumber\\
	\left\vert \Theta\right\rangle _{3}  &  =\left\vert 10\right\rangle =\left(
	\begin{array}
		[c]{cccc}%
		0 & 0 & 1 & 0
	\end{array}
	\right)  ^{T},\ \ \left\vert \Theta\right\rangle _{4}=\left\vert
	11\right\rangle =\left(
	\begin{array}
		[c]{cccc}%
		0 & 0 & 0 & 1
	\end{array}
	\right)  ^{T}\,. 
	\label{a.2}
\end{align}

A pure state $\left\vert \Psi\right\rangle _{AB}\in\mathcal{H}_{AB}$ is called
separable if and only it can be represented as $\left\vert \Psi\right\rangle
_{AB}=\left\vert \Psi\right\rangle _{A}\otimes\left\vert \Psi\right\rangle
_{B}$,\ $\left\vert \Psi\right\rangle _{A}\in\mathcal{H}_{A}$,$\ \left\vert
\Psi\right\rangle _{B}\in\mathcal{H}_{B}$. Respectively, a state $\left\vert
\Psi\right\rangle _{AB}$ is entangled if it is not separable.

A measure of the entanglement is a real positive number 
$E(\left\vert\Psi\right\rangle _{AB})\in\mathbb{R}_{+}$ which is assigned to each
state $\left\vert \Psi\right\rangle_{AB}$. The measure of entanglement is
zero for separable states, and assumes its maximum $1$ for maximally entangled states.

An entanglement measure $E(\left\vert\Psi\right\rangle_{AB})$
of a state $\left\vert \Psi\right\rangle _{AB}$ of a two-qubit system was
proposed by Bennett in Ref. \cite{Bennet}. It reads\footnote{We note that
	there exist also some different characteristics of entanglement measures, see
	Refs. \cite{9,11,12,13,14,15,16,17}.}:
\begin{align}
	&  E\left(  \left\vert \Psi\right\rangle _{AB}\right)  =S\left(  \hat{\rho
	}_{A}\right)  =S\left(  \hat{\rho}_{B}\right)  \ ,\nonumber\\
	&  S\left(  \hat{\rho}_{A}\right)  =-\mathrm{tr}\left(  \hat{\rho}_{A}\log
	\hat{\rho}_{A}\right)  ,\ \ S\left(  \hat{\rho}_{B}\right)  =-\mathrm{tr}
	\left(  \hat{\rho}_{B}\log\hat{\rho}_{B}\right)\,, 
	\label{a.3}
\end{align}
where $S(\hat{\rho}_{A})$ is von Neumann entropy of a statistical operator 
$\hat{\rho}_{A}$ of the subsystem $A$, whereas $S(\hat{\rho}_{B})$ is von 
Neumann entropy of the statistical operator $\hat{\rho}_{B}$ of the subsystem 
$B$ (one can see that $S(\rho_{A})=S(\rho_{B})$). For a pure state 
$\left\vert\Psi\right\rangle_{AB}$, we have:
\begin{align}
	& \hat{\rho}_{A}=\mathrm{tr}_{B}\hat{\rho}_{AB}=\sum_{b}\langle b\left\vert
	\hat{\rho}_{AB}\mid b\right\rangle\,,\nonumber\\
	& \hat{\rho}_{B}=\mathrm{tr}_{A}\hat{\rho}_{AB}=
	\sum_{a}\langle a\left\vert \hat{\rho}_{AB}\mid a\right\rangle,\quad
	\hat{\rho}_{AB}=\ _{AB}\mid\Psi\rangle\langle\Psi\mid_{AB}\,. 
	\nonumber
\end{align}

In the case of a pure state, its reduced statistical operators have nonzero
quantum entropy, whereas the entropy of the initial pure state is always zero.
By definition, a pure state of a two-qubit system is maximally entangled if
its reduced statistical operators are proportional to the identity operators.

Let us decompose the state $\left\vert \Psi\right\rangle_{AB}$ and the
operator $\hat{\rho}_{AB}$ in the computational bases,
\begin{align}
	&\left\vert \Psi\right\rangle _{AB}=\sum_{s=1}^{4}\upsilon_{s}\left\vert
	\Theta\right\rangle _{s}\Longrightarrow\nonumber\\
	&\hat{\rho}_{AB}=\left[  \upsilon
	_{1}\left\vert 00\right\rangle +\upsilon_{2}\left\vert 01\right\rangle
	+\upsilon_{3}\left\vert 10\right\rangle +\upsilon_{4}\left\vert
	11\right\rangle \right] \times\nonumber\\
	&\qquad\quad\left[  \upsilon_{1}^{\ast}\langle00\mid+\upsilon
	_{2}^{\ast}\langle01\mid+\upsilon_{3}^{\ast}\langle10\mid+\upsilon_{4}^{\ast}
	\langle11\mid\right]\,. 
	\nonumber
\end{align}
Then, with account taken of the relations
\begin{align}
	&  \left\vert 0\right\rangle \langle0\mid\ =\mathrm{diag}\left(  1,0\right),\qquad\,\,\,\,
	   \left\vert 1\right\rangle \langle1\mid\ =\mathrm{diag}\left(  0,1\right)\,,\nonumber\\
	&  \left\vert 0\right\rangle \langle1\mid\ =\mathrm{antidiag}\left(  1,0\right),\quad
	   \left\vert 1\right\rangle \langle0\mid=\mathrm{antidiag}\left(  0,1\right)\,,
	   \nonumber
\end{align}
where states $\left\vert a\right\rangle$ basis vectors in the Hilbert space
$\mathcal{H}_{A}$, we obtain:
\begin{align}
	&  \hat{\rho}_{A}=\ _{B}\langle0\mid\hat{\rho}_{AB}\left\vert 0\right\rangle
	_{B}+\ _{B}\langle1\mid\hat{\rho}_{AB}\left\vert 1\right\rangle _{B}
	=\nonumber\\
	& \mid\upsilon_{1}\mid^{2}\left\vert 0\right\rangle \langle0\mid+\upsilon_{1}
	\upsilon_{3}^{\ast}\left\vert 0\right\rangle \langle1\mid+\mid\upsilon_{2}
	\mid^{2}\left\vert 0\right\rangle \langle0\mid+\upsilon_{2}\upsilon_{4}^{\ast
	}\left\vert 0\right\rangle \langle1\mid\nonumber\\
	&  +\upsilon_{3}\upsilon_{1}^{\ast}\left\vert 1\right\rangle \langle
	0\mid+\mid\upsilon_{3}\mid^{2}\left\vert 1\right\rangle \langle1\mid+\upsilon_{4}
	\upsilon_{2}^{\ast}\left\vert 1\right\rangle \langle0\mid+\mid\upsilon_{4}
	\mid^{2}\left\vert 1\right\rangle \langle1\mid\ =\left(
	\begin{array}
		[c]{ll}
		\rho_{11}^{\left(  A\right)  } & \rho_{12}^{\left(  A\right)  }\\
		\rho_{21}^{\left(  A\right)  } & \rho_{22}^{\left(  A\right)  }
	\end{array}
	\right)  ,\nonumber\\
	&  \rho_{11}^{\left(  A\right)  }=\mid\upsilon_{1}\mid^{2}+\mid\upsilon_{2}\mid^{2}
	,\ \rho_{12}^{\left(  A\right)  }=\upsilon_{1}\upsilon_{3}^{\ast}+\upsilon
	_{2}\upsilon_{4}^{\ast}\ ,\nonumber\\
	& \rho_{21}^{\left(  A\right)  }=\upsilon
	_{3}\upsilon_{1}^{\ast}+\upsilon_{4}\upsilon_{2}^{\ast},\ \ \rho_{22}^{\left(
		A\right)  }=\mid\upsilon_{3}\mid^{2}+\mid\upsilon_{4}\mid^{2}\,. 
	\nonumber
\end{align}
Calculating the entanglement measure, we can use eigenvalues of the matrix
$\hat{\rho}_{A}$,
\begin{align}
	&  \hat{\rho}^{\left(  A\right)  }P_{a}=\lambda_{a}P_{a},\ \ \lambda_{a}
	=\frac{1}{2}\left[  \rho_{11}^{\left(  A\right)  }+\rho_{22}^{\left(
		A\right)  }+\left(  -1\right)  ^{a}y\right]  ,\ a=1,2;\nonumber\\
	&  P_{a}=\left(
	\begin{array}
		[c]{c}
		\frac{\rho_{11}^{\left(  A\right)  }-\rho_{22}^{\left(  A\right)  }+\left(
			-1\right)  ^{a}y}{2\rho_{21}^{\left(  A\right)  }}\\
		1
	\end{array}
	\right)  ,\ y=\sqrt[+]{\left(  \rho_{11}^{\left(  A\right)  }-\rho
		_{22}^{\left(  A\right)  }\right)  ^{2}+4\left\vert \rho_{12}^{\left(
			A\right)  }\right\vert ^{2}}\nonumber\\
	\  &  =\sqrt{\left(  \mid\upsilon_{1}\mid^{2}+\mid\upsilon_{2}\mid^{2}-\mid\upsilon_{4}
		\mid^{2}-\mid\upsilon_{3}\mid^{2}\right)  ^{2}+4\left\vert \upsilon_{1}\upsilon
		_{3}^{\ast}+\upsilon_{2}\upsilon_{4}^{\ast}\right\vert ^{2}}\,. 
	\label{s.8}
\end{align}
Thus, we obtain
\begin{equation}
	E\left(  \Psi\right)  =-\sum_{a=1,2}\lambda_{a}\log_{2}\lambda_{a}=-\left[
	z\log_{2}z+\left(  1-z\right)  \log_{2}(1-z)\right]  ,\ \ z=\frac{1+y}{2},
	\nonumber
\end{equation}
and by convention we adopt that $0\log_{2}0\equiv0$ (see, e.g., Ref. \cite{NieCh00}).

It is also known that for a pure two-qubit state one can recognize
entanglement by evaluating the so-called Schmidt measure $E_{\mathrm{S}}(\Psi)$, which is the trace of the squared reduced density operators,
\begin{align}
	& E_{\mathrm{S}}(\Psi)=1-\mathrm{tr}\left[  \left(  \hat{\rho}^{(1)}\right)
	^{2}\right]  =1-\sum_{a=1,2}^{2}\lambda_{a}^{2}=\nonumber\\
	& 1-\left[  \left(  \rho
	_{11}^{\left(  A\right)  }\right)  ^{2}+\left(  \rho_{22}^{\left(  A\right)
	}\right)  ^{2}+2\left\vert \rho_{12}^{\left(  A\right)  }\right\vert
	^{2}\right]\,. 
	\label{s.10}
\end{align}
The Schmidt measure can be considered as an alternative to the information
entanglement measure.

\end{document}